\documentclass{svjour3}
\usepackage{graphicx}
\usepackage{enumitem}
\usepackage{paralist}
\usepackage[normal,scriptsize]{subfigure}

\begin{document}
\title{Some peculiarities of motion of neutral and charged test particles
in the field of a spherically symmetric charged object in General
Relativity
}

\author{Valentin D. \,Gladush   \and
        Marina V. \,Galadgyi}


\institute{Valentin D. Gladush \at
              Dnepropetrovsk National University,\\
              72, Gagarin Ave, Dnepropetrovsk 49010, Ukraine \\
              \email{vgladush@dsu.dp.ua}
           \and
           Marina V. Galadgyi \at
              Dnepropetrovsk National University of Railway Transport,\\
              2, Lazarian St., Dnepropetrovsk, 49700, Ukraine\\
              \email{galadgyi@gmail.com}}

\date{Received: date / Accepted: date}
\maketitle
\begin{abstract}
We propose the method of investigation of radial motions for
charged and neutral test particles in the Reissner-Nordstr\"{o}m
field by means of mass potential. In this context we analyze
special features of interaction of charges and their motions in
General Relativity and  construct the radial motion
classification. For test particles and a central source with
charges $q$ and $Q$, respectively, the conditions of attraction
(when $qQ>0$) and repulsion (when $qQ<0$) are obtained. The
conditions of motionless test particle states with respect to the
central source are investigated and, in addition, stability
conditions for such static equilibrium states are found. It is
shown that stable states are possible only for the bound states of
weakly charged particles in the field of a naked singularity.
Frequencies of small oscillations of test particles near their
equilibrium positions are also found.

\end{abstract}

\section{Introduction}
\label{intro} Investigation of the test particle motion in General
Relativity is a classical research method of the structure and
properties of space-time near gravitating mass. If a
gravit\-ational field source is electrically charged, it
substantially influences the space-time geometry. Respectively,
the motion character of test bodies is changed. Therefore, the
investigation of motion peculiarities for charged particles in
such spaces is of a great physical interest as a part of general
research of  charged configurations in General Relativity.

In this paper we consider the test particles moving radially in
the gravitational field of a spherically symmetric source carrying
mass $M$ and charge $Q$. This field  is described by
Reissner-Nordstr\"{o}m metric
\begin{eqnarray}\label{metrika}
ds^{2}=Fc^{2}{d T}^{2}-F^{-1}{dR} ^2-R^{2}(d\theta^{2} +
\sin^{2}{\theta}{d\varphi}^{2})\,
\end{eqnarray}
where
\begin{eqnarray}\label{F}
F=1-\frac{2\gamma M}{c^{2}R}+ \frac{\gamma Q^{2}}{c^{4}R^{2}}\,,
\end{eqnarray}
$\gamma$ and $c$ are the gravitational constant and the velocity
of light respectively. The charge $Q$ also generates an electric
field with potential
\begin{equation}\label{pot}
    \varphi = \frac{Q}{R}\,.
\end{equation}
Note that one can differentiate the following 3 types of the
spherically symmetric charged relativistic objects depending on
the relation between their masses and charges: a charged black
hole (BH) ($\sqrt{\gamma} M>|Q|$)\,, an extremely charged black
hole (EBH) ($\sqrt{\gamma} M=|Q|$)\,, a the super-extremely
charged object ($\sqrt{\gamma}M<|Q|$ ---  naked singu\-larity)\,.

The\ standard\  qualitative\ \ analysis\ method\ for\ particle
dynamics\ uses the velocity\ potential\ $U_{V}$.\ This\ potential\
results\ from\ the\ radial\ \ equation\ of\ particle\  motion\
$\left(mc^{2}dR/ds \right)^{2}=$ $-U_{V}(M, Q, m, q, E, L, R)$
(see, for example,
\cite{dymnikova,piragas71,piragas95,cohen,gonsal,wilkins}) where
parameters $m$ and $q$ are\ particle\ mass\ and charge, $E$ and
$L$ are its total energy\ and angular orbital momentum in the
Reissner-Nordstr\"{o}m\ field\ with\ parameters\ $M$ and $Q$. The\
inequality\ $U_{V}(M, Q, m, q, E, L, R)\leq0$ defines the regions
of admissible radial motions of the particle. The solutions of the
equation $U_{V}(M, Q, m, q, E, L, R)=0$ with respect to $R$
specifies the turning points. Of course, this method is effective
for a complete system of the first integrals of the motion  equations.
In this case the radial motion equation of the first order is
possible.

In order to decrease the number of parameters $\{M, Q, m, q, E,
L\}$ and simplify the equations one can use a scale invariance of
the dynamical system 
(if such invariance is present) and the effective potential method
additionally. This potential is defined as a solution of the equation
$U_{V}(M, Q, m, q, E, L, R)=0$ with respect to one of parameters
$\{M, Q, m, q, E, L\}$. The\ solution\ is\ interpreted\ as
potential\ corresponding to the\ given\ parameter.\ The\
para\-me\-ter choice\ depends\ on\ the\ statement\ of the  problem.\
Thus,\ the\ mass\ potential\ $U_{M}=U_{M}(Q, m, q, E, L, R)$ is
determined as a solution of equation $U_{V}(U_{M}, Q, m, q, E, L,
R)=0$ with respect to $U_{M}$, the energy potential
$U_{E}=U_{E}(M, Q, m, q, L, R)$ is found as a solution of equation
$U_{V}(M, Q, m, q, U_{E}, L, R)=0$ with respect to $U_{E}$, and
the charge potential $U_{q}=U_{q}(M, Q, m, E, L, R)$ is determined
as a solution of equation  $U_{V}(M, Q, m, U_{q}, E, L, R)=0$
with respect to $U_{q}$, etc. It is easy to see that turning points
in the above cases are defined by relations $U_{M}=M$, $U_{E}=E$,
and $U_{q}=q$ respectively. Notice that the study of the radial
motion of charged test particles in the Reissner-Nordstr\"{o}m field
by means of energy potential $(U_{E})$ was carried out in
\cite{felice}. In fact, the methods of charged ($U_{q}$) and mass
($U_{M}$) potentials were used in \cite{cohen} and \cite{gladush},
respectively.

It turns out that the parameters included into the velocity
potential linearly lead to the simplest potentials. In this case,
the admissible motion region is defined by one inequality for a
new potential. For example, it is more convenient to use the mass
potential $U_{M}=U_{M}(Q, m, q, E, R)$, studying the one-dimensional
radial motion of the neutral and charged test particles ($L=0$) in
the field of a spherically symmetric charged source. This
potential application makes it possible to construct an evident
motion picture and gives a simple and natural classification of
particle trajectories.

The radial trajectories of particles in the Reissner-Nordstr\"{o}m
field were studied in \cite{cohen,felice,bicak,finley}. One can
also find the detailed motion description of neutral and charged
particles in the Reissner-Nordstr\"{o}m field in \cite{chandra}. The
special motion feature in the field of a charged source is sometimes the
fact that both neutral and charged particles with $qQ<0$ repel
from the gravitating center \cite{gron,cruz,leon,gorelik}.
In general, it takes place in the
case of charged shells and other matter distributions. This
phenomenon is directly associated with the stability of charged
relativistic configurations against a gravitational contraction
and with conditions of the gravitational collapse prevention.

Therefore, one of the main aims of such research is to obtain the
conditions when falling into the center is impossible (equilibrium
conditions for charged particles) and to find stability
conditions. For example, the paper \cite{bonnor} is devoted to such a
problem. However, the mentioned problem was not considered
there with the energy conservation law for a particle and so was
not solved completely. Stable, neutral and unstable
equilibrium conditions for charged particles in the field of a
gravitating center were discussed in \cite{cohen}. In our opinion,
those conditions where not obtained in the closed form because
they contain not only parameters of the particles and the central
source, but also the distance between them.

In the present paper, unlike the above mentioned
ones, we 
give the complete classification of the radial motions of test
particles in spherically symmetric gravitational and electric
fields of the central source. This classification naturally follows
the energy conservation law of particles and results from the
mass potential method and the method of the function of horizons
developed here for the first time. This paper is organized the
following way: section 2 contains the basic relations and the motion
equation of charged particles in the Reissner-Nordstr\"{o}m field.
In section 3 we analyze the special features of the test particle
motion and consider the cases of their exotic motion when
particles with charge $qQ>0$ attract to the center while those
with charge $qQ<0$ repel from it. In particular, here we
consider the radial oscillations of weakly charged particles in
the field of a super-extremely charged object and obtain their
oscillation period
in terms of proper time. Section 4 deals with stability
conditions. Here we find equilibrium conditions for particles in a
closed form and obtain both such state stability conditions and
equilibrium radius. These results  are expressed only in terms of
parameters of particles and the gravitating center $(M, Q, m, q,
E)$.

In this paper we neglect the radiation of charged particles and
the back reaction of radiation.  Note that the constructed
classification and stability conditions for stable states are
valid also in the case of test charged dust spherical shells, for
which the problem of radiation and its back reaction vanishes.


\section{Test particle motion equation}
\label{sec:1}

We take the action for a charged test particle with mass $m$ and
charge $q$ in the Reissner-Nordstr\"{o}m field in the form
\begin{equation}\label{action}
    S=-\int(mcds+q\varphi dT)=\int \mathcal{L}dT\,
\end{equation}
where $\mathcal{L}$ is the Lagrangian of the particle. In the case
of radial motions we have
\begin{eqnarray}\label{lagrangian}
\mathcal{L}=-mc\sqrt{Fc^{2}-F^{-1}{\dot{R}}^2} - qQ/R\,.
\end{eqnarray}
Here a dot denotes the derivative with respect to time $T$. The
action (\ref{action}) can be obtained, for example, through
generalizing an action of special relativity theory  for a charged
particle in an electromagnetic field \cite{landau} into the
corresponding action in General Relativity.

The total energy conservation law for a charged particle
\begin{eqnarray}\label{energy}
    E= \frac{mc^{2}F}{\sqrt{F-F^{-1}{\dot{R}}^2/c^{2}}} +
    \frac{q Q}{R} = mc^{3}F \,\frac{d T}{ds}+
    \frac{q Q}{R}={\mbox const}\,
\end{eqnarray}
implies the world line equation
\begin{equation}\label{eq/motion}
\left(mc^{2}\frac{dR}{ds}\right)^{2}=\left(E-\frac{q
Q}{R}\right)^{2}-m^{2}c^{4}\left(1-\frac{2\gamma
M}{c^{2}R}+\frac{\gamma Q^{2}}{c^{4}R^{2}}\right)\equiv - U_V(M,
Q, m, q, E, R)\,,
\end{equation}
\begin{equation}\label{dT/ds}
    mc^{2}\frac{dT}{ds}=
    \frac{1}{cF}\left(E-\frac{q Q}{R}\right)\,.
\end{equation}
Similar equations were obtained in
\cite{cohen,gonsal,felice,finley}. Here
\begin{eqnarray}\label{U}
U_V(M, Q, m, q, E, R)= m^{2}c^{4}-E^{2}-(\gamma m^{2}c^{2} M-E q
Q)\frac{2}{R}+ (\gamma m^{2}-q^{2})\frac{Q^{2}}{R^{2}}\
\end{eqnarray}
is  the velocity potential. \- Admissible \- motions are\- \-
determined \- by the inequality $U_V\leq 0\,$.\- The solutions of
the equation $U_V (M, Q, m, q, E, R) = 0 \, $ with respect to $R$
define the turning points. Using the equation (\ref{eq/motion})
for the particle acceleration we find
\begin{eqnarray}\label{acceleration}
\frac{d^{2}R}{ds^{2}}=\frac{1}{m^{2}c^{4}}\left[ \left( E q
Q-\gamma m^{2}c^{2}M \right) \frac{1}{R^{2}} + \left(\gamma
m^{2}-q^{2}\right)\frac{Q^{2}}{R^{3}}\right]\,.
\end{eqnarray}
The given relations yield the trajectory classification of
radially moving particles. Further, one can find static
equilibrium positions of particles and  examine their stability.

\section{Radial motion classification}
\label{sec:2}

The object of our classification is a central source with
parameters $\{M, Q \}$ and a test particle with parameters $\{m,
q, E\}$ moving in the field of the source. It is inconvenient to
classify motions by means of the velocity potential $U_V (M, Q, m,
q, E, R) $  depending on all five parameters $\{M, Q, m, q, E, R
\}$. Note that the Lagrangian (\ref{lagrangian}), energy
(\ref{energy}), and the motion equation (\ref {eq/motion}) are
invariant with respect to scale transformation
\begin{equation}\label{inv/cond}
  (M, Q, m, q,  E, R, T, s)\longrightarrow (aM, aQ, am, aq,  aE, aR, aT, as)\,.
\end{equation}
Therefore, we fix one of the parameters, for example, $Q\neq 0$\,,
and, to be definite, assume that $Q>0$ and $M>0$.

The subsequent simplification and decrease of the number of
parameters is achieved by introducing a new potential. Since the
parameter $M$ is included into the velocity potential (\ref{U})
linearly, it is convenient to use the mass potential $U_{M}$
defined by
\begin{equation}\label{UM}
U_V(U_{M}, Q, m, q, E, R)= m^{2}c^{4}-E^{2}-(\gamma m^{2}c^{2}
U_{M}-E q Q)\frac{2}{R}+ (\gamma
m^{2}-q^{2})\frac{Q^{2}}{R^{2}}=0.
\end{equation}
Hence, for the mass potential we obtain
\begin{equation}\label{Um}
   U_{M}= \frac{1}{2\gamma m^{2}c^{2}}
 \left[(m^{2}c^{4}-E^{2})R+ 2Eq Q  +
\left(\gamma m^{2}-q^{2}\right)\frac{Q^{2}}{R}\right] \,.
\end{equation}
It follows from (\ref{U}) that
\begin{equation}\label{intrUM}
    U_{M}-M=\frac{R}{2\gamma m^{2}c^{2}}\,U_V\,.
\end{equation}
The condition \ $U_{V}\leq 0\,$ gives the inequality \ $U_{M}(Q,
m, q, E, R)\leq M \,$  defining the admissible regions of
particle motions  in terms of the mass potential. The equation
solutions $U_{M}(Q, m, q, E, R) = M$ with respect to $R$ specify
the turning points.

Note that the potential $U_{M}$ depends only on the particle
parameters and has the form of a linear-fractional function, in
addition. Therefore, the mass potential properties are  
determined by its asymptotical behavior at $R\rightarrow 0$ and at
$R\rightarrow\infty$, i.e. by coefficients at $R$ and at $1/R$.

It turns out that the mass potential behavior $U_{M}$ when
$R\rightarrow 0$ depends on the electrical characteristics of the
particle:

 \quad 1) $U_{M}\rightarrow +\infty$ if
  $\gamma m^{2}> q^{2}$  (the weakly charged particle),%

 \quad 2)  $U_{M}\rightarrow Eq Q/\gamma m^{2}c^{2}$, if
 $\gamma m^{2}= q^{2}$ (the extremely charged particle),%

 \quad 3) $U_{M}\rightarrow -\infty$ if  $\gamma m^{2}< q^{2}$
(the super-extremely charged particle);\\
while the behavior of $U_{M}$ when $R\rightarrow \infty $ depends
only on the energy characteristics of the particle: 

\quad 1) $U_{M}\rightarrow +\infty$ if
$E^{2}<m^{2}c^{4}$  (the bound states of the particle),

\quad 2) $U_{M}\rightarrow Eq Q/\gamma m^{2}c^{2}$,
if  $E^{2}=m^{2}c^{4}$ (the particle with a critical mass),

\quad 3)  $U_{M}\rightarrow  -\infty$, if $E^{2}>m^{2}c^{4}$ (the
unbound states of the  particle).

In coordinates $\{R, U\}$ the plot of the mass potential at the
fixed parameters $\{Q, m, q, E\}$ is represented by a curve
$U=U_{M}(Q, m, q, E, R)$. The regions of admissible motions are
determined by the horizontal segments of straight lines
$U=M=const\,$ lying above the curve $U=U_{M}(Q, m, q, E, R)$.
Intersection points of the curve $U=U_{M}(Q, m, q, E, R)$ and the
line $U=M=const$ give the turning radii.

The complete history of particles can be observed on the Penrose
diagram for the extended Reissner-Nordstr\"{o}m space-time. This
problem was studied in
\cite{bicak,finley,chandra,cruz,vikers,graves} and, therefore, is
not considered here. Nevertheless,  we plot the curves of horizons
for the Reissner-Nordstr\"{o}m metric to have some information
about the space-time structure. With this aim in view, we
introduce the additional function of horizons $U_{h}=U_{h}(Q, R)$
as a solution of the equation $F(U_{h}, Q, R) =1-{2 \gamma
U_{h}}/{c^{2} R} + {\gamma Q^{2}} / {c^{4} R^{2}} =0 $ with
respect to $U_{h}$:
\begin{eqnarray}\label{Uh}
U_{h}=U_{h}(Q, R) = \frac{1}{2}\left(\frac{Rc^{2}}
{\gamma}+\frac{Q^{2}}{Rc^{2}}\right)\,.
\end{eqnarray}
If the charge $Q$ is fixed, the function of horizons $U_{h}(Q, R)$
determines the BH mass with the horizon radius $R$. It is easy to
see that $U_{h}\geq|Q|/\sqrt{\gamma}$ and we have the minimum
$({U}_{h})_{min}=|Q|/\sqrt{\gamma}=M$ for the EBH at
$R=|Q|\sqrt{\gamma}/c^2$. In  coordinates $\{R, U\}$,  the
intersection points of the curve $U=U_{h}(Q, R)$ and the straight
line $U=M=const$ give the radii of horizons $R_{\pm}$ which the
particle passes through.

Comparing functions $U_{M}(Q,m,q,E,R)$ and $U_{h}(Q, R)$, we get
the relation
\begin{equation}\label{UhUM}
    U_{h}(Q, R)=U_{M}(Q,m,q,E,R)+\frac{1}{2\gamma m^2 c^2}
    \left(E\sqrt{R}-\frac{qQ}{\sqrt{R}}\right)^{2}\,.
\end{equation}
It follows from (\ref{UhUM}) that
\begin{equation}\label{uneq}
    U_{h}(Q, R) \geq U_{M}(Q,m,q,E,R)\,.
\end{equation}
This means that in coordinates $\{R, U\}$ the curve $U=U_{h}(Q,
R)$ lies always above the curve $U=U_{M}(Q, m, q, E, R)\,$. The
intersection points of the straight line $U=M =const$ with
horizons curve $U=U_{h}(Q, R)$ are in the region of admissible
motions $U_{M}\leq M \,$. The latter corresponds to the case when
the turning radii cannot be in T-region ($R_{-}<R<R_{+}$ where
$R_{\pm} =c^{-2}(\gamma M \pm \sqrt{\gamma^{2}M^{2} - \gamma
Q^{2}}\,)$ are exterior and interior Reissner-Nordstr\"{o}m
horizons) where the radial coordinate becomes time-like. For
neutral particles with $E=0$ the potential $U_{M}(Q, m, q, E, R)$
coincides with the function ${U}_{h}(Q, R)$. E The equality
$U_{h}(Q, R)= U_{M}(Q, m, q, E, R)\,$ at $E\neq 0 \,,$ $q \neq0$
defines a tangency point of these curves: $R_{t} =qQ/E$. The same
point for a particle with energy $E=qQ/R_{+}$ is a turning point
on event horizon \cite {chandra}  with $R _ {t} =R_{+}$. Thus, the
particle never passes into R-regions ($R<R_{-}$ and $R_{+}<R$). If
function $U_{M}$ has a minimum, then we have $(U_{M})_{min}\leq
({U}_{h})_{min} = |Q |/ \sqrt {\gamma}$ from the inequality
(\ref{uneq}).

Let us consider the particle acceleration. It follows from
(\ref{acceleration}) that neutral particle acceleration vanishes
at $R=Q^{2}/Mc^2$. Thus, in the region $R>Q^{2}/Mc^2$  attraction
takes place, whereas repulsion occurs in the region
$R<Q^{2}/Mc^2$.

For positively charged particles the acceleration vanishes at
\begin{eqnarray}\label{Ra(accel=0)}
R=R_{a}\equiv\frac{(\gamma m^2 - q^{2})Q^{2}}{\gamma m^{2}c^{2} M
- E q Q}\,.
\end{eqnarray}
From the condition $R_{a}>0$ and (\ref{acceleration}) it is clear
that attraction takes place at $R>R_{a}$ for weakly charged
particles with $0 <q<\sqrt{\gamma}m\,$  and $\gamma
m^{2}c^{2}M>EqQ$\,.  Attraction is observed also at $0<R<R_{a}$
for super-extremely charged particles with $q>\sqrt{\gamma}m\,$
and $\gamma m^{2}c^{2}M<EqQ$. For super-extremely charged
particles with $\gamma m^{2}c^{2} M \geq EqQ$ attraction occurs
for all $0<R<\infty$.

For extremely charged particles the acceleration sign depends only
on the central source and particle parameters
\begin{equation}\label{acsel}
    \frac{d^{2}R}{ds^{2}}=
    \frac{\gamma}{c^4 R^2}\left(E\,\frac{Q}{q}-Mc^2\right)\,.
\end{equation}
If $E/q <Mc^2/Q $, attraction can take place, but $E/q> Mc^2/Q $
results in repulsion. When the energy-to-charge ratio of a
particle $E/q$ is equal to  the ratio of the ``total
energy"-to-charge of the central object $Mc^2/Q$ (i.e. $E/q =
Mc^2/Q$), the gravitation and electric interactions are completely
compensated and ${d^{2}R}/{ds^{2}} =0$. Extremely charged
particles with a critical mass also move with the vanishing
acceleration in the field of an EBH.

For negatively charged particles repulsion is possible only for
weakly charged particles  $-\sqrt{\gamma}m <q<0\,$ in the region
\begin{eqnarray}\label{Ra(accel1=0)}
0<R<\tilde{R}_{a}\equiv\frac{(\gamma m^2 - q^{2})Q^{2}}{\gamma
m^{2}c^{2} M +E |q| Q}\,.
\end{eqnarray}

Let us introduce the dimensionless quantities
\begin{eqnarray}\label{undim}
    \left.\begin{array}{l}
     \mathcal{U}_{M}=\sqrt{\gamma}U_{M}/Q\,, \quad
    \mathcal{U}_{h}=\sqrt{\gamma}U_{h}/Q\,,\quad
    x=R c^{2}/\sqrt{\gamma }Q\,,\\
     \varepsilon=E/mc^{2}\,,\qquad
    \beta = q/\sqrt{\gamma}m\,,\qquad \mathcal{M}= \sqrt{\gamma}M/Q\,.
  \end{array}  \right.
\end{eqnarray}
Then  the expressions (\ref{Um}) and (\ref{Uh}) can be rewritten
as
\begin{eqnarray}
&& \mathcal{U}_{M}(\varepsilon, \beta,
x)=\frac{1}{2}\left[(1-\varepsilon^{2})x + 2\varepsilon \beta
+\left(1-\beta^{2}\right)
\frac{1}{x}\right]\,,\label{Um/undim}\\
&& \qquad \qquad \mathcal{U}_{h}(x)= \frac{1}{2}\left( x +
\frac{1}{x}\right)\,.\label{Uh/undim}
\end{eqnarray}
Function graphs $\mathcal{U}_{M}(x)$ and $\mathcal{U}_{h}(x)$ в in
dimensionless coordinates $\{ x\,,\mathcal{U} \}$ are displayed in
Fig. 1-3 in solid and dashed lines, respectively.

\begin{figure}[p!]
\begin{center}$
\begin{array}{ccc}
\subfigure[Bound states of particles ($E^{2}<m^{2}c^{4}$).]
{\includegraphics[width=3.5cm,height=3.5cm]{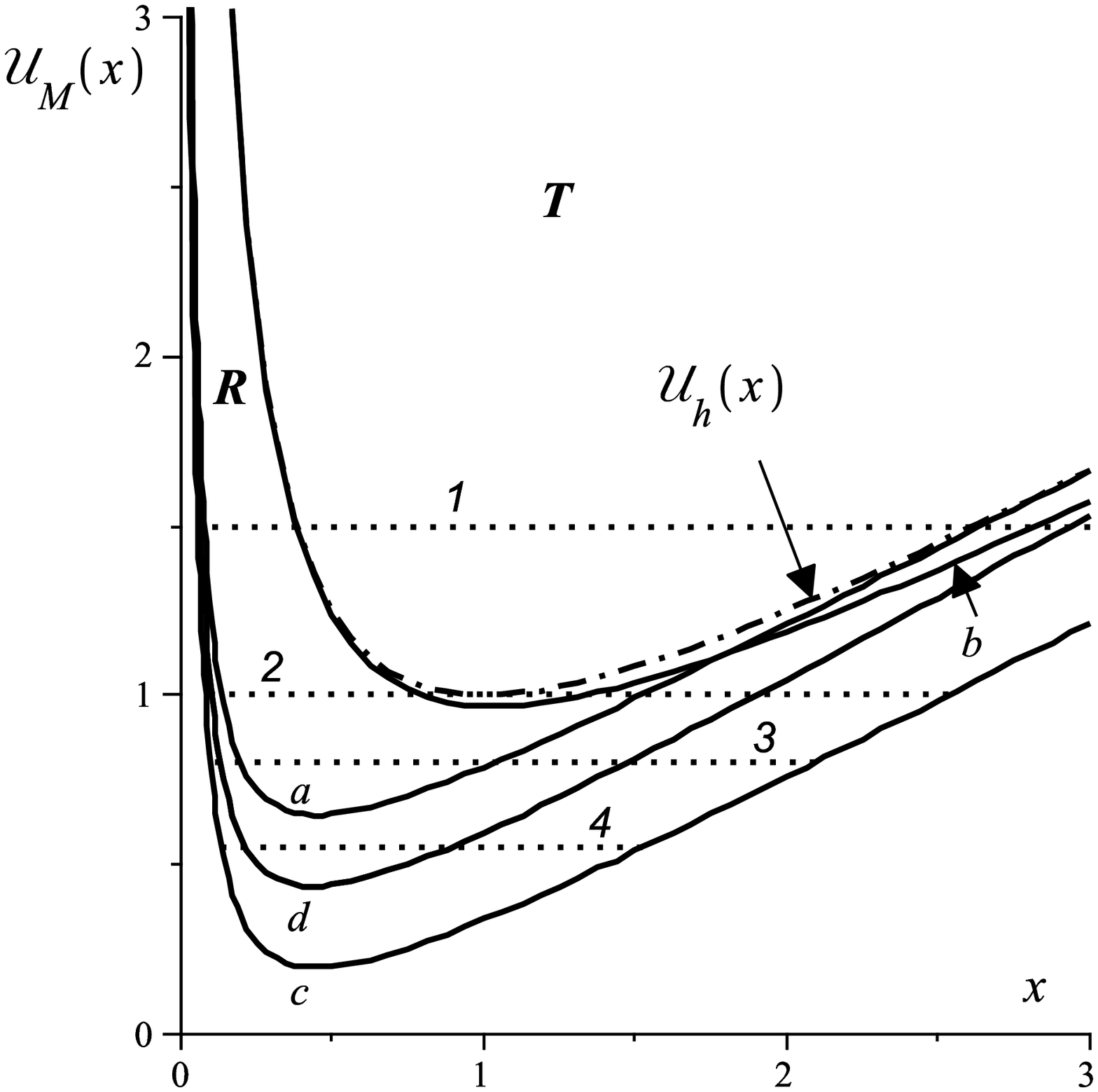}\label{w-a}}~~
& \subfigure[Particles with a cri\-ti\-cal mass
($E^{2}=m^{2}c^{4}$).]
{\includegraphics[width=3.5cm,height=3.5cm]{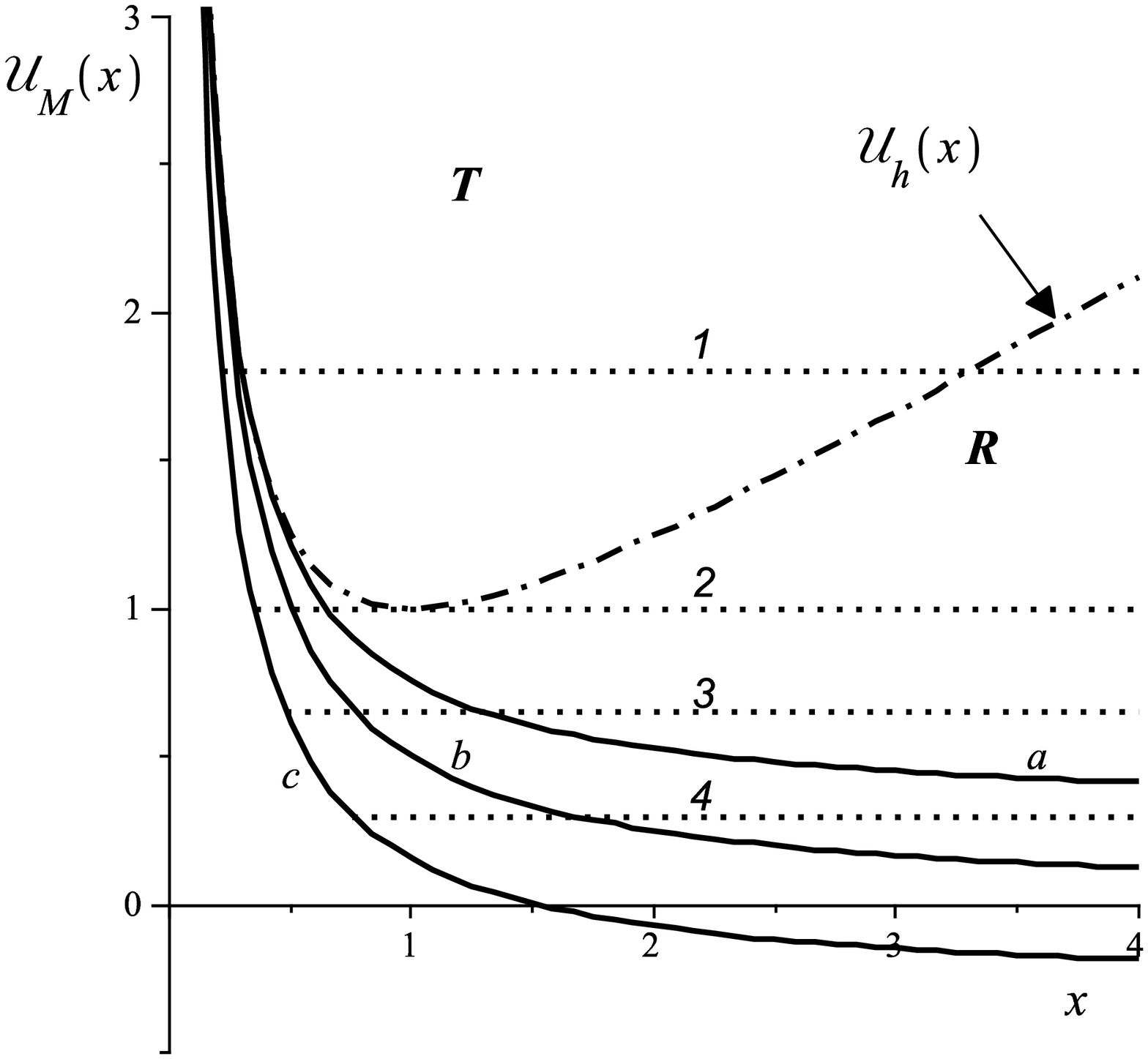}\label{w-b}}~~
& \subfigure[Unbound states of particles ($E^{2}>m^{2}c^{4}$).]
{\includegraphics[width=3.5cm,height=3.5cm]{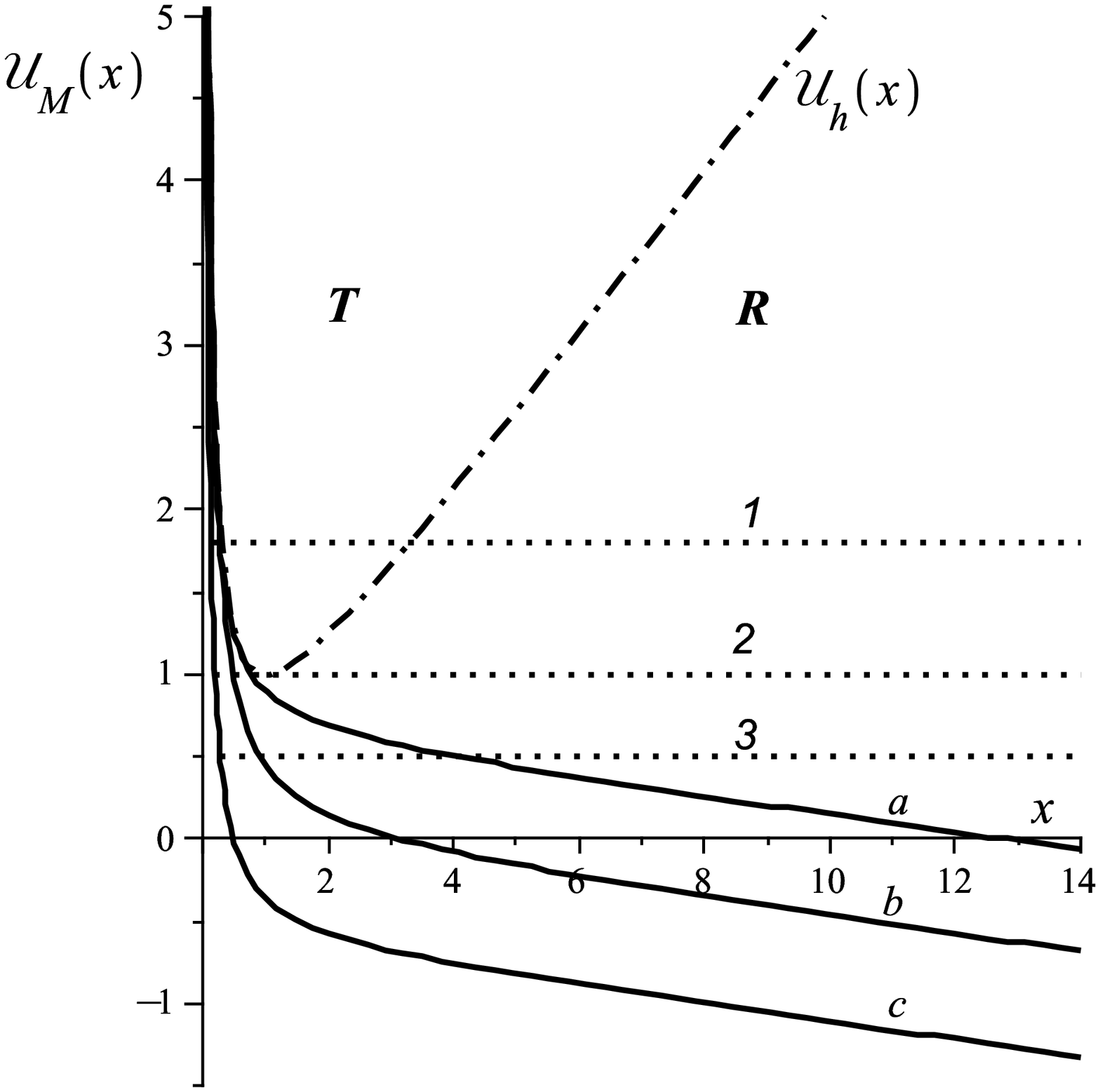}\label{w-c}}
\end{array}$
\end{center}
\caption{Weakly charged particles ($\gamma
m^{2}>q^{2}$).}\label{w}
\end{figure}
\begin{figure}[p!]
\begin{center}$
\begin{array}{ccc}
\subfigure[Bound states of particles ($E^{2}<m^{2}c^{4}$).]
{\includegraphics[width=3.5cm,height=3.5cm]{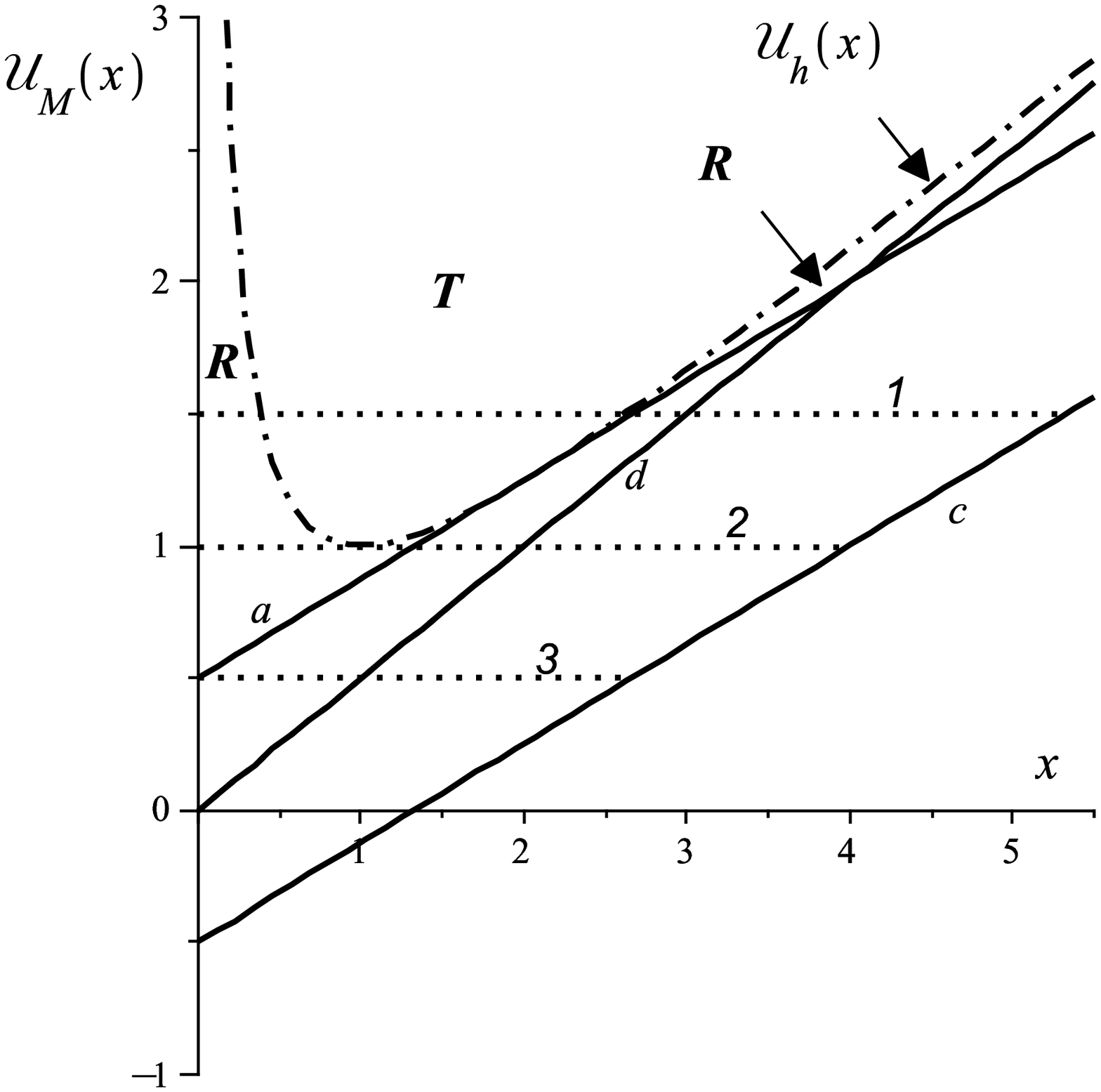}\label{ex-a}}~~&
\subfigure[Particles with a cri\-ti\-cal mass
($E^{2}=m^{2}c^{4}$).]
{\includegraphics[width=3.5cm,height=3.5cm]{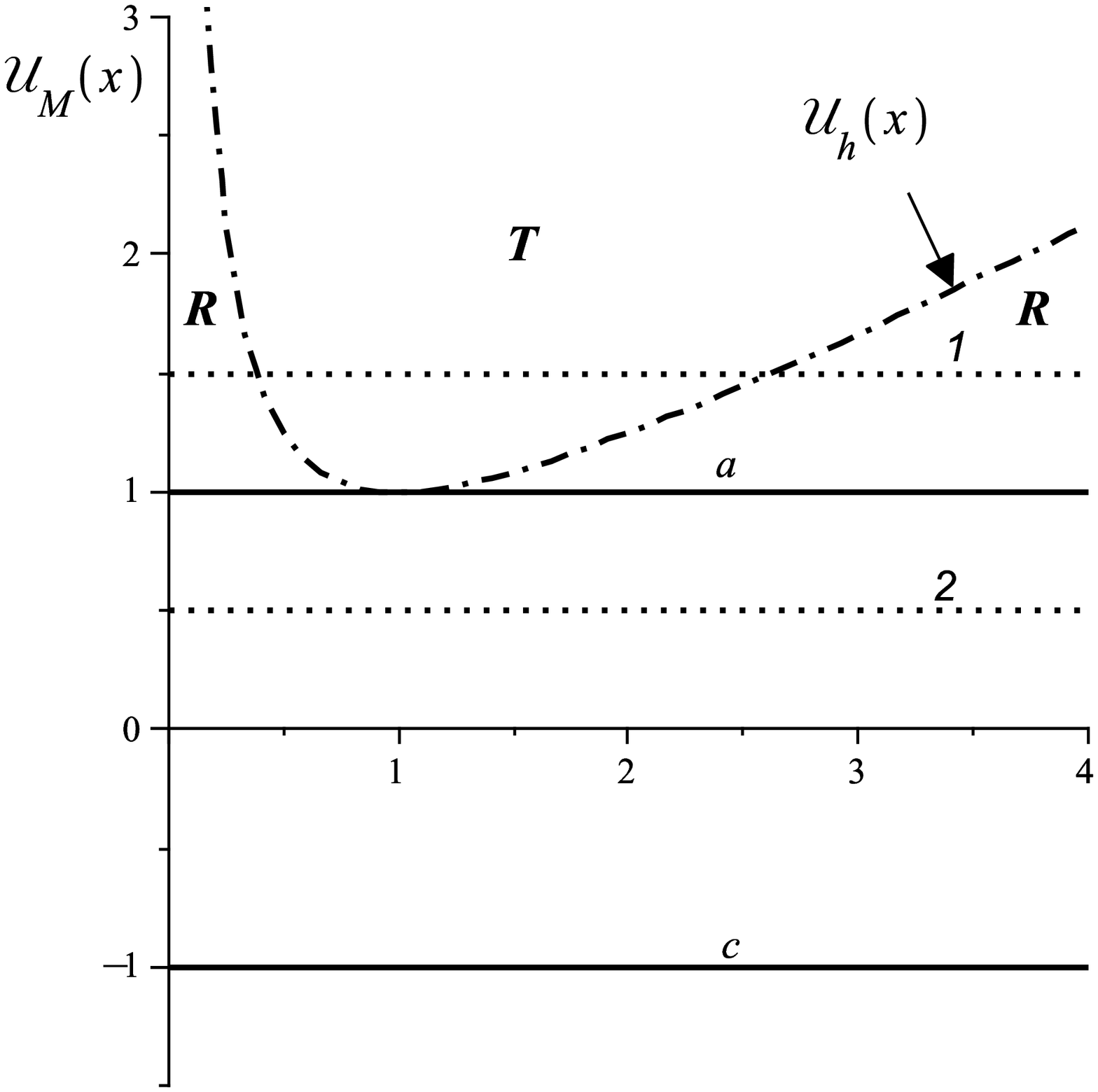}\label{ex-b}}~~&
\subfigure[Unbound states of particles ($E^{2}>m^{2}c^{4}$).]
{\includegraphics[width=3.5cm,height=3.5cm]{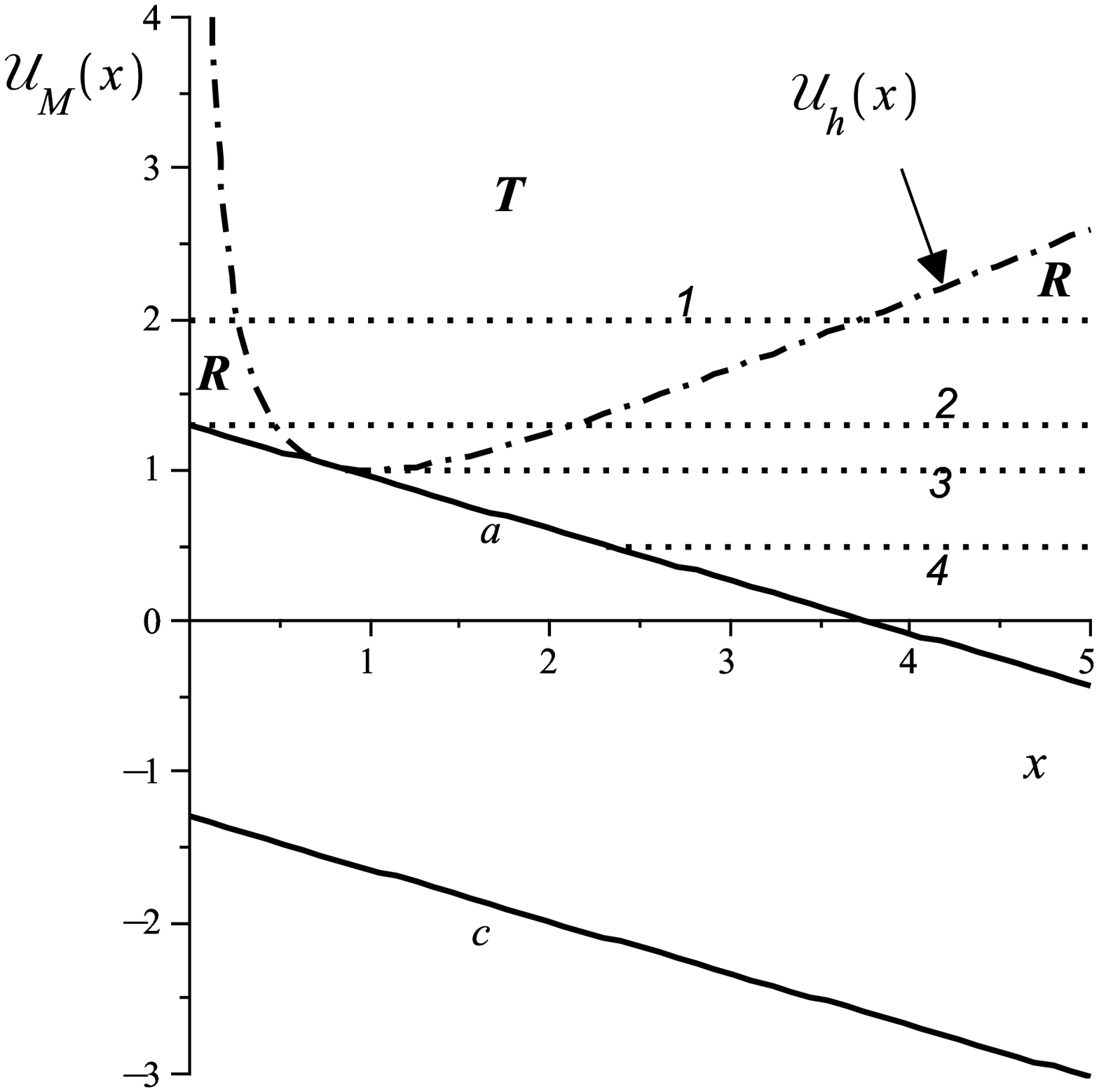}\label{ex-c}}
\end{array}$
\end{center}
\caption{Extremely charged particles ($\gamma
m^{2}=q^{2}$).}\label{ex}
\end{figure}
\begin{figure}[p!]
\begin{center}$
\begin{array}{ccc}
\subfigure[Bound states of particles ($E^{2}<m^{2}c^{4}$).]
{\includegraphics[width=3.5cm,height=3.5cm]{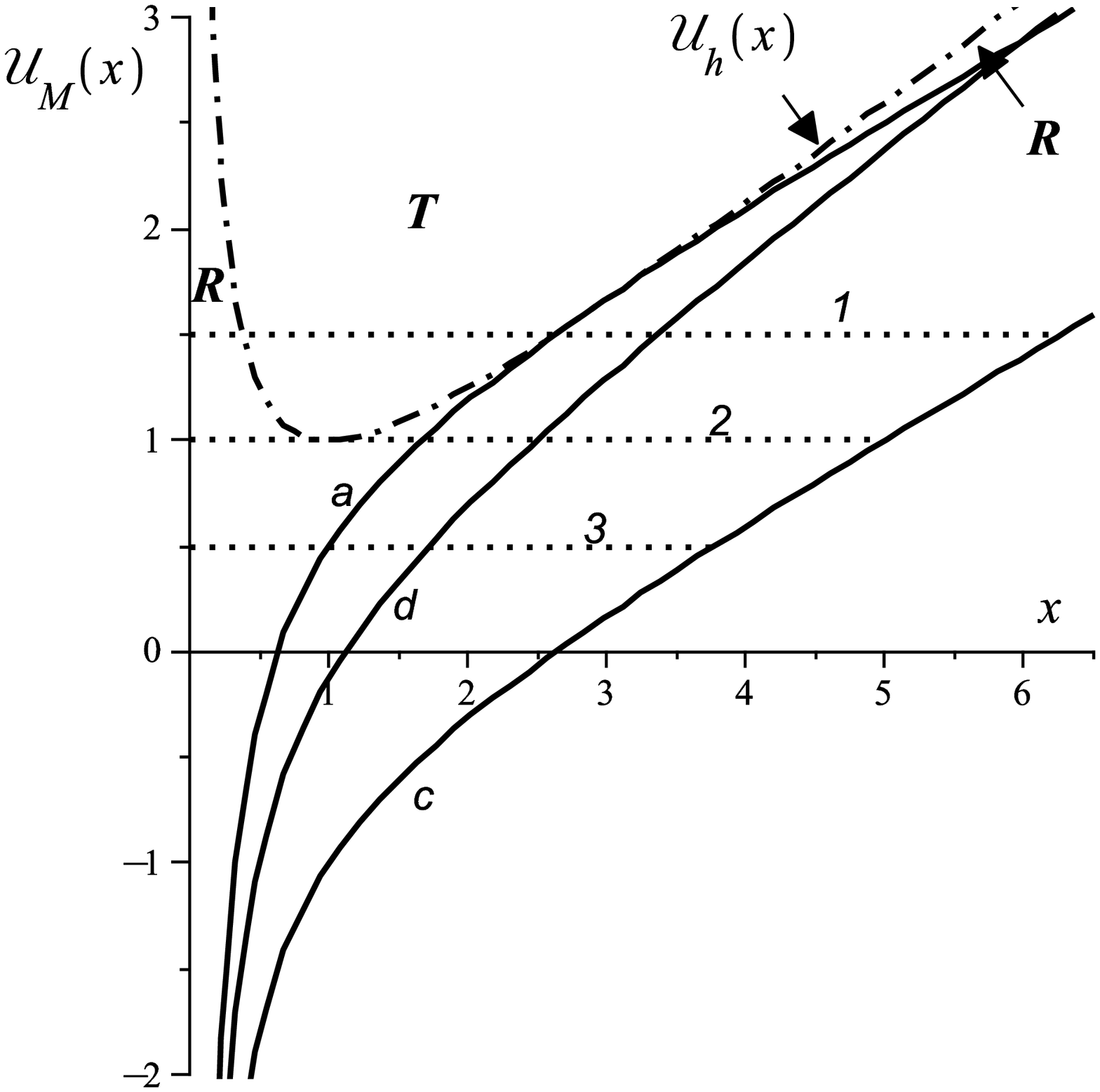}\label{ab-a}}~~&
\subfigure[Particles with a cri\-ti\-cal mass
($E^{2}=m^{2}c^{4}$).]
{\includegraphics[width=3.5cm,height=3.5cm]{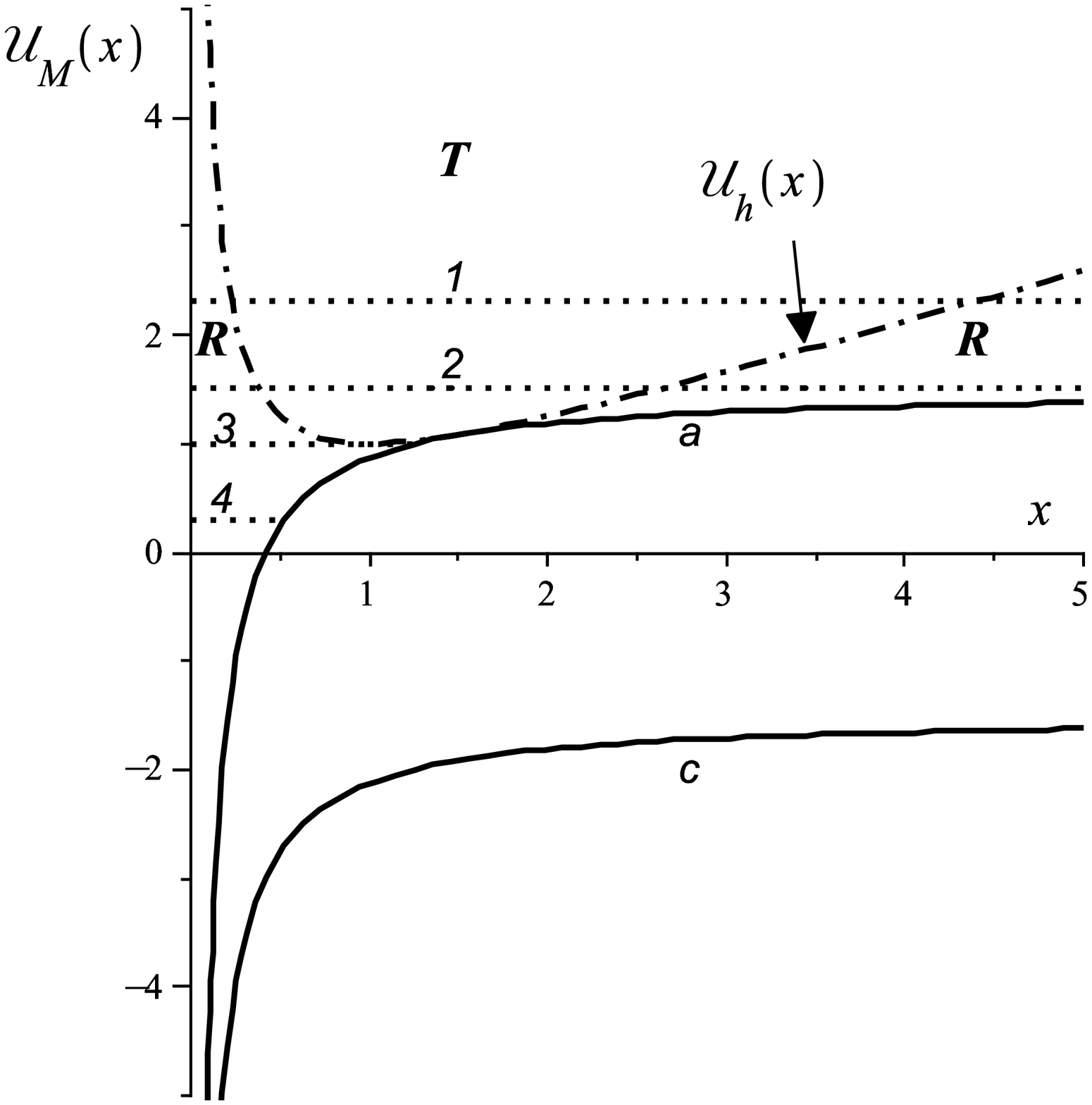}\label{ab-b}}~~&
\subfigure[Unbound states of particles ($E^{2}>m^{2}c^{4}$).]
{\includegraphics[width=3.5cm,height=3.5cm]{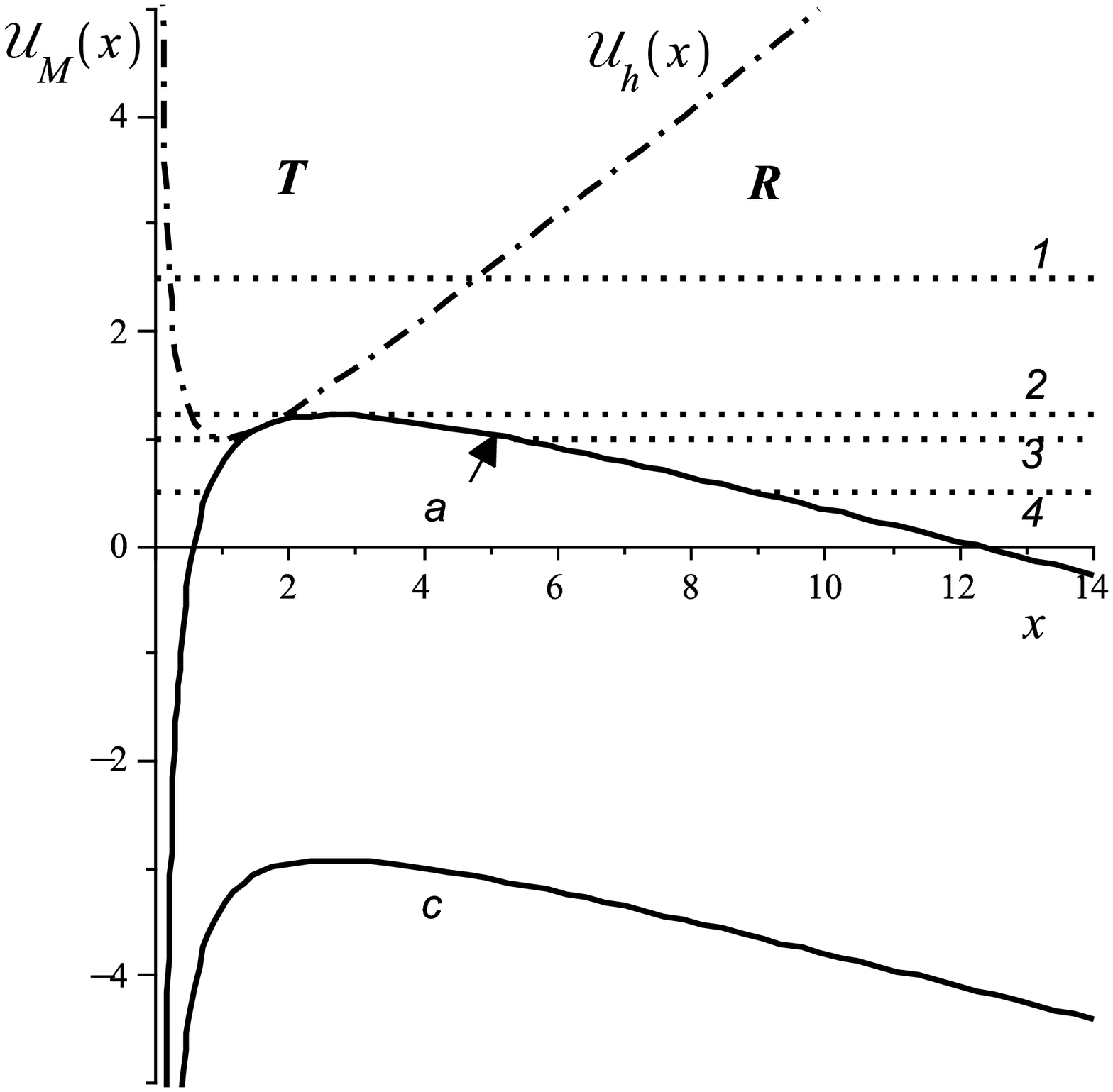}\label{ab-c}}
\end{array}$
\end{center}
\caption{Super-extremely charged particles ($\gamma m^{2}<
q^{2}$).}\label{ab}

{\scriptsize{ In\- figures\- $1$, $2$\- and\- $3$ the\-
admissible\- motion\- regions\- for\- given\- $\mathcal{M}$ are
determined by the horizontal segments of dotted \-lines\-
$\mathcal{U}=\mathcal{M}=const$ lying above the curve
$\mathcal{U}= \mathcal{U}_{M}(\varepsilon, \beta, x)$. The
segments of lines $\mathcal{U}=\mathcal{M}>1$,
$\mathcal{U}=\mathcal{M}=1$ and $\mathcal{U}=\mathcal{M}<1$
describe particle motions in the field of a BH, EBH, and
super-extremely charged object (naked singularity), respectively.
The intersection points of the mass potential curve  $\mathcal{U}=
\mathcal{U}_{M}(\varepsilon, \beta, x)$ and straight line
$\mathcal{U}=\mathcal{M}=const$ give the turning radii. The curves
$a$, $b$ and $c$ correspond to mass potentials for the cases
$q>0$, $q=0$ and $q<0$, respectively. The curve  $d$ corresponds
to the  mass potential of a particle with energy $E=0$}}.
\end{figure}

It is possible to specify basic particle motion types depending on
the relations between parameters  $\{m, q, E\}$ and  the central
source mass $M$\,:

\textbf{1.} \textbf{Weakly charged particles:} $\gamma m^{2}>
q^{2}$ or $\beta^{2}<1$ (Fig. \ref{w}). We can differentiate
several motion cases with regard to energy parameters.

\textit{1.1.} \textit{Bound states of weakly charged particles:}
$E^{2}<m^{2}c^{4}$ or $\varepsilon^{2}<1$. The particles move
inside the potential well between the turning radii $R_{1}$ and
$R_{2}$ (Fig. \ref{w-a}) where
\begin{eqnarray}\label{1.1(R1_R2)}
R_{1, 2}=\frac{\gamma M m^{2}c^{2}-EqQ\mp m
\sqrt{\triangle}}{m^{2}c^{4}-E^{2}}\,,
\end{eqnarray}
here
\begin{eqnarray}\label{delta(R1_R2)}
        \triangle
 =\gamma(E Q-qMc^2)^2+c^4(\gamma M^2-Q^2)(\gamma m^2-q^2)
 \,.
\end{eqnarray}
Hence one can see that $\triangle >0$ when $Q\leq\sqrt{\gamma}M$.
If $Q>\sqrt{\gamma}M$, the requirement $\triangle>0$ gives the
energy condition either $E\geq E_{+}$ or $E\leq E_{-}$ for the
particle where
\begin{eqnarray}\label{energpm}
E_{\pm}  =\frac{c^{2}}{\sqrt{\gamma}}
\left(\frac{q}{Q}M\sqrt{\gamma}\pm\sqrt{\gamma
m^2-q^2}\sqrt{1-\frac{\gamma M^{2}}{Q^{2}}}\,\right)\,.
\end{eqnarray}
The central source mass is bounded by the inequality
$(U_{M})_{min}\leq M<\infty$ where
\begin{equation}\label{U_min(1.1)}
(U_{M})_{min}=U_{M}(R_{extr})=\frac{Q}{\gamma m^{2}c^{2}}\left(Eq
+ \sqrt{(m^{2}c^{4}-E^{2})(\gamma
m^{2}-q^{2})\,}\,\right)<\frac{Q}{\sqrt{\gamma}}\,.
\end{equation}
If $M=(U_{M})_{min}$, the particle is at the bottom of the
potential well. Thus, relations $dR/ds=0$ and $d^{2}R/ds^{2}=0$
are satisfied, and the particle with energy $E=E_{+}$ remains
motionless at the distance
\begin{eqnarray}\label{R_extr(1.1)}
R_{extr}=Q \sqrt{\frac{\gamma m^{2}-q^{2}}{m^{2}c^{4}-E_{+}^{2}}}\,
\end{eqnarray}
from the super-extremely charged central source. Taking into
account (\ref{energpm}), this formula can be rewritten as
\begin{equation}
R_{extr}=\frac{\sqrt{\gamma }Q^{2}}{c^{2}}\,\frac{\sqrt{\gamma m^{2}-q^{2}}}{%
\sqrt{\gamma }M\sqrt{\gamma m^{2}-q^{2}}-qQ\sqrt{1-\gamma
M^{2}/Q^{2}}}.
\end{equation}

If $M>(U_{M})_{min}$, the equation (\ref{eq/motion}) yields the
motion trajectory $R=R(s)$ in an implicit form
\begin{eqnarray}\label{R(s)}
\begin{array}{l}{\displaystyle \qquad\qquad s(R)-s_{0}=
\frac{2mc^{2}(\gamma M m^{2}c^{2}-E q Q)}
{(m^{2}c^{4}-E^{2})^{3/2}}
\arctan \sqrt{\frac{R-R_{1}}{R_{2}-R}}\ - }\\ \\
  {\displaystyle -\frac{mc^{2}}{m^{2}c^{4}-E^{2}}
  \sqrt{(E^{2}-m^{2}c^{4})R^{2}+2(\gamma M m^{2}c^{2}- E q Q)R
   -Q^{2}(\gamma  m^{2}-q^{2})}}\,.
\end{array}
\end{eqnarray}
In this case the particle oscillates anharmonically with the
period
\begin{eqnarray}\label{T}
T=\frac{2}{c}(s(R_{2})-s(R_{1}))=2\pi mc
 \frac{\gamma M m^{2}c^{2}-E q Q}{(m^{2}c^{4}-E^{2})^{3/2}}\,
\end{eqnarray}
with respect to the proper time. Thus, if $Q>\sqrt{\gamma}M$, then
the particle oscillates near the equilibrium position $R_{extr}$
in the field of the super-extremely charged object. In the case of
small deviations of the charged particle from the equilibrium
position we have harmonic oscillations with the period
\begin{equation}
T=\frac{2\pi \gamma mQ^{3}}{c^{3}}\frac{\sqrt{\gamma m^{2}-q^{2}}}{%
\left( M\sqrt{\gamma}\sqrt{\gamma m^{2}-q^{2}}-q\sqrt{Q^{2}-\gamma M^{2}}%
\right)^{2}}.
\end{equation}
If $Q<\sqrt{\gamma}M$, the region II $(R_{-}\leq R \leq R_{+})\,$
of the Penrose diagram for the maximally extended
Reissner-Nordstr\"{o}m solution enters the admissible motion
region $(R_{1}\leq R \leq R_{2})$. The particle trajectory
starting in the given region I crosses the horizon $R_{+}$, enters
the region II, and then crosses the horizon $R_{-}$. After that,
the particle reaches the turning point $R=R_{1}$ in the region
III, returns back, and passes through a new  region II into
another asymptotically flat region I of the Penrose diagram (see,
for example, \cite{hawking}, Fig. 25). Here the particle reaches
the turning point $R_{2}$ and comes back to the horizon $R_{+}$
again, etc. Therefore, the particle moves along the infinite chain
of regions I, II, III and horizons $R_{\pm}$: $... \rightarrow $ I
$\rightarrow R_{+}\rightarrow $ II $\rightarrow R_{-} \rightarrow
$ III $\rightarrow R_{-} \rightarrow $ II $\rightarrow R_{+}
\rightarrow $ I $\rightarrow...$.

In the case of EBH $(Q=\sqrt{\gamma} M )$ the particle moves along
the infinite chain of regions I, III and horizons $R_{e}=\gamma
M/c^{2}$: $... \rightarrow $ I $\rightarrow R_{e} \rightarrow $
III $\rightarrow R_{e} \rightarrow $ I $ \rightarrow...$  (see
\cite{hawking}, Fig. 26).

The bound state of a neutral particle with $E^{2}<m^{2}c^{4}$ and
$q=0$ is a special case of the bound state of a weakly charged
particle. The particle is in the potential well (Fig. \ref {w-a},
curve $b$) and moves within the region
\begin{eqnarray}\label{1.1(0R1_R2)}
R_{01}=\frac{\gamma M m^{2}c^{2}- m
\sqrt{\triangle}}{m^{2}c^{4}-E^{2}}\leq R \leq R_{02}=
\frac{\gamma M m^{2}c^{2}+ m \sqrt{\triangle}}{m^{2}c^{4}-E^{2}}\,
\end{eqnarray}
where $\triangle=\gamma^{2} M^{2} m^{2}c^{4}- \gamma Q^{2}(
m^{2}c^{4}-E^{2})$. The central source mass varies in the interval
$(U_{M})_{min}\leq M<\infty$ where
\begin{equation}\label{U_min(01.1)}
(U_{M})_{min}=U_{M}(R_{extr})=\frac{Q}{\sqrt{\gamma}}\,
\sqrt{1-\frac{E^{2}}{mc^{2}}\,}\, <\frac{Q}{\sqrt{\gamma}}\,.
\end{equation}
In the case when $M = (U_{M})_{min}$ the neutral particle with energy
\begin{eqnarray}\label{0ener}
   E=mc^{2}\sqrt{1- {\gamma M^{2}}/{Q^{2}}} \,
\end{eqnarray}
is at the bottom of the potential well and remains motionless at
the distance
\begin{eqnarray}\label{R0_extr(1.1)}
R_{extr0}=\frac{\sqrt{\gamma}\, m Q}{\sqrt{ m^{2}c^{4}-E^{2}}}=\frac{Q^{2}}{Mc^{2}} \,
\end{eqnarray}
from the super-extremely charged object. In the case when
$M>(U_{M})_{min}$ the particle oscillates anharmonically and we
have
\begin{eqnarray}\label{R0(s)}
 s(R)-s_{0}=
\frac{2\gamma m^{3}c^{4} M} {{(m^{2}c^{4}-E^{2})}^{{3}/{2}}}
 \arctan \sqrt{\frac{R-R_{01}}{R_{02}-R}}\
-mc^{2}\sqrt{\frac{(R-R_{01})(R_{02}-R)}{m^{2}c^{4}-E^{2}}}\,
\end{eqnarray}
with the period  $T=2\pi \gamma m^{3}c^{3}M/
(m^{2}c^{4}-E^{2})^{3/2}$ with respect to the proper time.

If $Q>\sqrt{\gamma}M$, the particle oscillates near the point
$R_{extr0}$ in the naked singularity field. In the case of small
deviations of the neutral particle from the equilibrium position
$R_{extr0}$ we have harmonic oscillations with the period
\begin{equation}
T=\frac{2\pi Q^{3}}{M^{2}c^{3}\sqrt{\gamma}}\,
\end{equation}
which is particle mass independent, as we see.

If $Q <\sqrt{\gamma}M$, the  particle starting in a given region I
passes through regions II, III and II and appears in another
asymptotically flat region I on the Penrose diagram (see
\cite{hawking}, Fig. 25).

In the case when $q=0$ and $E=0$ functions ${U}_{h}(Q, R)$ and
$U_{M}(Q,m,q,E,R)$ coincide that results in equality
$R_{01,02}=R_{\pm}$ and the particle moves along the world line
$$s(R)-s_{0}= {2\gamma  M}/{c^{2}}
\arctan \sqrt{\frac{R-R_{-}}{R_{+}-R}}\ -
 R\sqrt{-F}\,,$$
traveling along the infinite chain of regions II, sequentially
arriving to horizons $R_{+}$ and $R_{-}$: $... \rightarrow $ II
$\rightarrow R_{-} \rightarrow $ II $ \rightarrow R_{+}
\rightarrow $ II $ \rightarrow...$ (see. \cite{hawking}, Fig. 25).
Thus the particle moves from a horizon $R_{+}$ to the nearest
following horizon $R_{+}$ during the proper time $T=2\pi\gamma
M/c^{3}$.

\textit{1.2. Weakly charged particles with a critical mass}:
$E^{2}=m^{2}c^{4}$ or $\varepsilon^{2}=1$ (Fig. \ref{w-b}). The
mass potential has a simple form
\begin{equation}\label{Um12}
   U_{M}(Q,m,q,R)= \frac{1}{2\gamma m^{2}c^{2}}
 \left[2m q Q c^2  +
\left(\gamma m^{2}-q^{2}\right)\frac{Q^{2}}{R}\right] \,.
\end{equation}

The central object mass is bounded by inequality $qQ/\gamma m \leq
M<\infty$. For all masses $M> qQ/\gamma m$ the motion region is
bounded by the condition
\begin{eqnarray}\label{1.2_R_3}
  R\geq R_{3}=
\frac{Q^{2}(\gamma m^{2}-q^{2})}{2mc^{2}(\gamma m M-q Q)}\,.
\end{eqnarray}
The  particles start falling from the infinity with zero initial
velocity, reach the turning radius $R= R_{3}$, and eventually come
back to infinity again. In the case when $\gamma mM=qQ$
the particles are at rest at $R=+\infty$.%

For a neutral particle with a critical mass the motion region is
bounded by the condition $R\geq R_{03}=Q^{2}/2Mc^{2}$.

\textit{1.3.  Unbound states of weakly charged particles}:
$E^{2}>m^{2}c^{4}$ or $\varepsilon^{2}>1$ (Fig. \ref{w-c}). In
this case motion occurs in the region
\begin{equation}\label{1.3_R_4}
R\geq R_{4}=\frac{E q Q- \gamma M m^{2}c^{2} + m
\sqrt{\triangle}}{E^{2}-m^{2}c^{4}}\,.
\end{equation}
The particles start moving from the infinity with the initial
velocity $dR/ds=-\sqrt{E^2/m^2c^4 -1}$. Then they reach the
turning radius $R= R_{4}$ and go back to the infinity again.

For neutral particles in unbound states the motion region is
bounded by the condition $R\geq R_{04}$ where
\begin{equation}\label{1.3_R04}
 R_{04}=\frac{- \gamma M m^{2}c^{2} + m
\sqrt{\gamma^{2} M^{2} m^{2}c^{4}+ \gamma Q^{2}(
E^{2}-m^{2}c^{4})}}{E^{2}-m^{2}c^{4}}\,.
\end{equation}

\textbf{2. Extremely charged particles:}
 $\gamma m^{2}= q^{2}$ or $\beta^{2}=1$ (Fig.\ref{ex}). %
In this case the mass potential depends on $R$ linearly
\begin{eqnarray}\label{U_m_2}
U_{M}(Q,q,E,R)=\frac{1}{2\gamma q^2 c^{2}}
\left[(q^{2}c^{4}-\gamma E^{2})R+ 2\gamma q QE\right]\,.
\end{eqnarray}

\textit{2.1. Bound states of extremely charged particles:}
$E^{2}<m^{2}c^{4}$ or $\varepsilon^{2}<1$ (Fig.\ref{ex-a}). The
central object mass is bounded by the inequality $M\geq
EQ/\sqrt{\gamma}mc^{2}$. If $M>EQ/\sqrt{\gamma}mc^{2}$, the
particles move within the region $0\leq R\leq R_{5}$ (see dashed
lines $1$, $2$ and $3$ for all particles except particles with
$q>0$ for straight line $3$). Here
\begin{eqnarray}\label{2.1_R_5}
R_{5}=\frac{2\sqrt{\gamma} m(\sqrt{\gamma}mMc^{2}+E
Q)}{m^{2}c^{4}-E^{2}}\,.
\end{eqnarray}

In the case $M= EQ/\sqrt{\gamma}mc^{2}$ the particle will always
stay
at singularity $R=0$ %

\textit{2.2. Extremely charged particles with a critical mass:}
$E^{2}=m^{2}c^{4}$ or $\varepsilon^{2}=1$ (Fig. \ref{ex-b}). The
mass potential is independent of the radius $U_{M}(Q,m,q)=mQ/q$.
We have for the velocity and acceleration of particles
\begin{eqnarray*}
    \left(\frac{dR}{ds}\right)^{2}=\left( M
 -  \frac{q}{|q|}\frac{Q}{\sqrt{\gamma}}\right)
   \frac{2\gamma}{Rc^2}\,, \qquad
\frac{d^{2}R}{ds^{2}}=-
  \left(M - \frac{q}{|q|}\,\frac{Q}{\sqrt{\gamma}} \right)
  \frac{\gamma}{ R^2c^2}<0\,
\end{eqnarray*}
The central source mass is bounded by the inequality $M\geq
qQ/|q|\sqrt{\gamma}$. In the case when $q>0$ and $M >
Q/\sqrt{\gamma}\,$ (BH) or $q<0$ and $M>0$ particles move in the
region $0\leq R \leq \infty$. Here attraction takes place. If
$Q=M\sqrt{\gamma}\,$ (EBH), both the velocity and the acceleration
of particles with $q>0$ equal zero. The particle are in the
neutral equilibrium state and are at rest at arbitrary $R$.

\textit{2.3. Unbound states of extremely charged particles:}
$E^{2}>m^{2}c^{4}$ or $\varepsilon^{2}>1$ (Fig. \ref{ex-c}). In
the case when particles have a charge $q>0$, there are two
possibilities. If the central source mass varies within the
interval $0<M< EQ/\sqrt{\gamma}mc^{2}$, the particle acceleration
is positive $d^{2}R/ds^{2}>0$ and repulsion takes place. The
motion occurs in the region $R\geq R_{6}$ (straight lines $3$ and
$4$) where
\begin{eqnarray}\label{2.3_R_6}
R_{6}=\frac{2\sqrt{\gamma}m(E Q-\sqrt{\gamma} M
mc^{2})}{E^{2}-m^{2}c^{4}}\,.
\end{eqnarray}
Therefore, a particle coming from the infinity reaches $R_{6}$ and
goes back to the infinity again. If $M\geq
EQ/\sqrt{\gamma}mc^{2}$, the particle acceleration  is negative
$d^{2}R/ds^{2}<0$ and attraction takes place. Particles coming
from the infinity reach singularity $R=0$ (straight lines $1$ and
$2$). In the case $q<0$ particles fall to singularity for all
$M>0$.

\textbf{3. Super-extremely charged particles:} $\gamma m^{2}<
q^{2}$ or $\beta^{2}>1$ (Fig. \ref{ab}).%

\textit{3.1. Bound states of super-extremely charged particles:}
$E^{2}<m^{2}c^{4}$ or $\varepsilon^{2}<1$. Particles move in the
bounded region $0\leq R\leq R_{7}$ for all $M>0$ (Fig. \ref
{ab-a}, straight lines $1$, $2$ and $3$) where
 \begin{eqnarray}\label{3.1_R_7}
R_{7}=\frac{\gamma M m^{2}c^{2}-E q Q+ m
\sqrt{\triangle}}{m^{2}c^{4}-E^{2}}\,.
\end{eqnarray}

\textit{3.2. Super-extremely charged particles with a critical
mass:} $E^{2}=m^{2}c^{4}$ or $\varepsilon^{2}=1$ (Fig.
\ref{ab-b}). The mass potential has the form (\ref{Um12}). If the
central source mass is in the interval $0<M <qQ / \gamma m $,
particles always move within the region $0\leq R\leq R_{8}$
(straight lines $3 $ and $4 $) where
\begin{eqnarray}\label{3.2_R_8}
R_{8}=\frac{(q^{2}-\gamma m^{2})Q^{2}}{2 m c^{2}(q Q-\gamma M
m)}\,.
\end{eqnarray}
If $M \geq qQ / \gamma m $, all particles falling from the
infinity reach singularity $R=0$ (straight lines $1 $ and $2 $). %

\textit{3.3. Unbound states of super-extremely  charged
particles:} $E^{2}>m^{2}c^{4}$ or $\varepsilon^{2}>1$ (Fig.
\ref{ab-c}). The mass potential  is bounded above
$U_{M}(Q,R)\leq(U_{M})_{max}$ where
\begin{equation}\label{U_max_3.3}
(U_{M})_{max}=U_{M}(Q, \tilde{R}_{extr})=\frac{Q}{\gamma
m^{2}c^{2}}\left(Eq-\sqrt{(E^{2}-m^{2}c^{4})(q^{2}-\gamma m^{2})}\right)\,,
\end{equation}
\begin{eqnarray}\label{R_extr_for max}
\tilde{R}_{extr}=Q \sqrt{\frac{q^{2}-\gamma
m^{2}}{E^{2}-m^{2}c^{4}}}\,.
\end{eqnarray}
If $M>(U_{M})_{max}$ (straight line 1), a particle with $q>0$
coming from the infinity inevitably falls to singularity. If
$M=(U_{M})_{max}$ (straight line 2), a particle falling from the
infinity reaches $R_{extr} $. At this point the conditions
$dR/ds=0$ and $d^{2}R/ds^{2}=0$ are satisfied and the maximum of
the potential corresponds to the unstable equilibrium position. If
$M <(U_{M})_{max}$ (straight line $3 $ and $4 $), the particle
motion occurs either in the region $0\leq R\leq R_{9}$, or $R\geq
R_{10}$, where the particle reaches singularity. Here
\begin{eqnarray}\label{3.3_R_9,10}
R_{9, 10}=\frac{E q Q- \gamma Mm^{2}c^{2}\mp
m\sqrt{\bigtriangleup}}{E^{2}-m^{2}c^{4}}\,.
\end{eqnarray}
A particle with $q <0 $ traveling from the infinity reaches
singularity $R=0$ for all  $M>0 $.

\section{Stability conditions}
\label{sec:4}

Stationary equilibrium states of a particle at some $R=R_{extr}$
are defined by conditions $d R/d s = 0$ и $d^{2}R/d s^{2}=0$.
Furthermore, if $d^{2} R/d s^{2}>0$ at $R< R_{extr}$ and $d^{2}
R/d s^{2}< 0$ at $R> R_{extr}$, there exist a stable position
$R_{extr}$. It follows from definitions for the velocity and mass
potentials (\ref{intrUM}) that at the potential well bottom (Fig.
\ref{w-a}) in the cases $U(R_{extr})=0$ or $U_{M}(R_{extr})=M$ the
conditions $(\partial U/\partial R)|_{R_{extr}} = 0$ and
$(\partial U_{M}/\partial R)|_{R_{extr}} = 0$ coincide and give a
static equilibrium radius $R_{extr}$ (\ref{R_extr(1.1)}).

Taking into account (\ref{eq/motion}) and (\ref{acceleration}), we
can eliminate a variable $R$ from the equations $d R/d s=0$ and
$d^{2}R/d s^{2}=0$. As a result we obtain the equation
$\triangle=0$ (see (\ref{delta(R1_R2)})) which can be rewritten as
\begin{eqnarray}\label{Rex}
(m^{2}c^{4}-E^{2})(\gamma m^{2}-q^{2})Q^{2}=({m^{2}c^{2}\gamma M
-E q Q})^{2}\,.
\end{eqnarray}
From  here one can find two systems of inequalities
\begin{equation}\label{sys1a}
|q| < m \sqrt{\gamma},\quad |E|< mc^{2}\,
\end{equation}
or
\begin{equation}\label{sys1b}
|q| > m \sqrt{\gamma},\quad |E|> mc^{2} \,.
\end{equation}
Rewriting (\ref{Rex}) in the form
\begin{eqnarray}\label{Rex1}
(Q^{2}-\gamma M^{2})(\gamma m^{2}-q^{2})=\gamma \left(\frac{E
Q}{c^{2}}-qM\right)^{2}\,,
\end{eqnarray}
we obtain in a similar way
\begin{equation}\label{sys2a}
|q| < m \sqrt{\gamma},\quad |Q|> M \sqrt{\gamma}\,\\
\end{equation}
or
\begin{equation}\label{sys2b}
|q| > m \sqrt{\gamma},\quad |Q|< M \sqrt{\gamma}\,.
\end{equation}
Therefore, from the motion classification and  relations
(\ref{sys1a}) and (\ref{sys2a}) we find the following conditions:
$|E|< mc^{2}$, ~$|q|< m \sqrt{\gamma}$ and ~$|Q|> M
\sqrt{\gamma}$\,. This case corresponds to the stable equilibrium
of the bound states of a weakly charged particle in the field of a
super-extremely charged object \footnote{ It is noteworthy that in
\cite {frolov} it is approved otherwise: the stable system is
impossible in the case of a super-extremely charged object $|Q |>
M \sqrt {\gamma}$.}. The relations (\ref{sys1b}) and (\ref{sys2b})
yield the following conditions: $|E|> mc^{2}$, ~$|q|> m
\sqrt{\gamma}$ and ~$|Q|< M \sqrt{\gamma}$\,. This case
corresponds to the instable equilibrium of the unbound states of a
super-extremely charged particle in the field of a charged BH. The
stability conditions (\ref{sys2a}) and (\ref{sys2b}) have been
obtained in \cite{bonnor}, but there was not pointed out which of
the equilibrium states was stable, and a static equilibrium radius
has not been found for the particle.

We find from (\ref{Rex1}) that a charged test particle at the
static position $R_{extr}$ (\ref{R_extr(1.1)}) has energy $E =
E_{+}$ (see (\ref{energpm})). In turn, a neutral test particle
remains motionless at the distance $R_{extr0}$
(\ref{R0_extr(1.1)}) and has  energy (\ref{0ener}).

\section{Conclusion}
\label{conclus}

A special feature of the charge interaction in General Relativity
is the fact that in the case of test particles with charge $q>0$
moving radially in the field of a central source with charge $Q>0$
both  for weakly charged particles $(q<m\sqrt{\gamma})$ at
$R>R_{a}$ and  for super-extremely charged particles
$(q>m\sqrt{\gamma})$ at $0<R<R_{a}$ attraction is observed (see
(\ref{Ra(accel=0)})). If $\gamma m^{2}c^{2}M\geq EqQ$, attraction
takes place for super-extremely charged particles for all values
$0<R<\infty$.

In the case of extremely charged particles $q=m\sqrt{\gamma}$ the
acceleration sign is independent of the distance; it depends only
on the parameters of the central source and particles. If $E/q <
Mc^2/Q$, attraction can take place, but $E/q > Mc^2/Q$ results in
repulsion. If the equality $E/q = Mc^2/Q$ is satisfied,
gravitational and electric interactions are  balanced completely.

In the case $q<0$ repulsion is possible only for weakly charged
particles $-m\sqrt{\gamma}<q<0\,$ for radius values
$0<R<\tilde{R}_{a}$ (see formula (\ref{Ra(accel1=0)})).

It is interesting to note that for neutral particles in the
Reissner-Nordstr\"{o}m field attraction takes place when
$R>Q^{2}/Mc^2$ whereas repulsion occurs when $R<Q^{2}/Mc^2$.

Introducing the mass potential made it possible to construct a
simple and evident classification of possible states for the
interacting system under consideration: a test particle with
parameters $\{m,E,q\}$ and a central source with parameters
$\{M,Q\}$. As a result, it has been shown that a stable static
position was possible only for bound states $(E^{2}< m^{2}c^{4})$
of weakly charged particles $(\gamma m^{2}>q^{2})$ moving in the
field of a charged naked singularity $(\gamma M^{2}<Q^{2})$. In
this case the central object mass is equal to the mass potential
minimum $M=(U_{M})_{min}$ (see (\ref{U_min(1.1)})). Incidentally
energy $E_{+}$ and static position $R_{extr}$ for a particle are
determined by formulae (\ref{energpm}) and (\ref{R_extr(1.1)}). If
$M>(U_{M})_{min}$, the particle oscillates near its equilibrium
position $R_{extr}$. Let us note that the motionless state is also
observed for neutral particles. In this case a static position
$R_{extr0}$ and energy $E$ are defined by formulae (\ref{0ener})
and (\ref{R0_extr(1.1)}).

It is easy to see that the equilibrium equations (\ref {Rex}) and
(\ref {Rex1}) are invariant with respect to the scale
transformation $M=aM'$, $Q=aQ'$, $E=aE'$, $m=am'$ and $q=aq'$.
Hence, according to (\ref{R_extr(1.1)}), $R_{extr}=aR_{extr}'$ and
we have the similarity law. If  we increase the charge and mass of
the central source and, correspondingly, the mass, charge and
energy of the particle by a factor $a$, the static state of the
particle with new parameters will remain stable, and its static
position radius will be also increased by the factor $a$.

The obtained stability conditions for the static state of a
particle can be generalized both for the case of static states of
charged dust shells  \cite{glad-galad} and  the case of charged
dust layers in spherically symmetric configurations in General
Relativity \cite{gladush07}. These conditions make it possible to
construct classical particle-like models of stable static charged
dust balls in General Relativity \cite{gladush08}.

The existence of the stable lowest state for a neutral or weakly
charged particle in the field of a charged naked singularity
points out also the existence of stationary quantum states of a
charged or neutral particle with a discrete spectrum of energies
in the field of a super-extremely charged central source of a
Reissner-Nordstr\"{o}m type. Thus, according to section $3$, in
the case of small deviations of a weakly charged particle from the
stable positions in the field of a naked singularity harmonic
oscillations take place. The oscillation frequency has the form
\begin{equation}
\omega =\frac{2\pi }{T}=\frac{c^{3}}{\gamma mQ^{3}}\frac{\left( M\sqrt{%
\gamma }\sqrt{\gamma m^{2}-q^{2}}-q\sqrt{Q^{2}-\gamma M^{2}}\right) ^{2}}{%
\sqrt{\gamma m^{2}-q^{2}}}\,.
\end{equation}
If we formally consider a quantum oscillator corresponding to this
case,  the lowest state of such system will have energy
$E_{0}=\hbar \omega /2$. For a neutral particle we will have
\begin{equation}
E_{0}=\frac{\hbar}{2}\frac{c^3M^2\sqrt{\gamma}}{Q^3}\,.
\end{equation}
Thus the lowest energy state of a neutral particle depend on the
naked singularity parameters $M$ and $Q$.

\begin{acknowledgements}
This work was supported  by the grant of the "Cosmomicrophysics"
program of the Physics and Astronomy Division of the National
Academy of Sciences of Ukraine.
\end{acknowledgements}


\begin{thebibliography}{99}
%

\def\CMPh{{\it Commun. Math. Phys.}~}
\def\JPh{{\it J. Phys.}~}
\def\CJP{{\it Czech.  J.  Phys.}~}
\def\LMPh {{\it Lett. Math. Phys.}~}
\def\NPh  {{\it Nucl. Phys.}~}
\def\PhE  {{\it Phys. Essays}~}
\def\PhL  {{\it Phys. Lett.}~}
\def\PhR  {{\it Phys. Rev.}~}
\def\PhRL {{\it Phys. Rev. Lett.}~}
\def\PhRp {{\it Phys. Rep.}~}
\def\NCim {{\it Nuovo Cimento}~}
\def\NuPB {{\it Nucl. Phys.}~}
\def\GRG {{\it Gen. Relativ. Gravit.}~}
\def\CQG {{\it Class. Quantum Grav.}~}
\def\prp {report}
\def\Prp {Report}

\def\GrC {{\it Gravitation $\&$ Cosmology}~}
\def\DANS {{\it Dokl. Akad. Nauk SSSR}~}
\def\APh {{\it Ann. Phys.}~}
\def\JMM {{\it Journ. Math. and Mech.}~}
\def\JMP {{\it J. Math. Phys.}~}
\def\IVUZ {{\it Izv. Vyssh. Uchebn. Zaved. Fiz}~}
\def\APP {{\it Acta Phys. Pol.}~}
\def\UFZh {{\it  Ukr. Fiz. Zh.}~}
\def\TMF {{\it Teor. Mat. Phys.}~}
\def\ZEMF {{\it Zh. Eksp. Teor. Fiz.}~}
\def\SPhJ {{\it Sov. Phys. J.}~}
\def\SPhJETP {{\it Sov. Phys. JETP.}~}
\def\PhEs {{\it Phys. Essays}~}
\def\RMP {{\it Rev. Mod. Phys}~}
\def\IJMD {{\it Int. Journ. Mod. Phys.}~}


\bibitem{dymnikova} Dymnikova, I.G.: Motion of particles and photons in the gravitational field of a rotating body,
\textit{Uspekhi Fiz. Nauk} \textbf{148}, 393-432 (1986).

\bibitem{piragas71}  Piragas, A.K., Krivenko, O.P.: Some questions
of qualitative theory for geodesics in the field of gravitating
center,  26 p. Institute for Theoretical Physics of the Ukrainian
Academy of Sciences (Preprint Ukraine Acad. Scien. ITP-71-113P),
Kiev (1971).

\bibitem{piragas95}  Piragas, A.K.,  Akeksandrov, A.N.  et al.:
Qualitative and analytic methods in relativistic dynamics, 448 p.
Energoatomizdat, Minsk (1995).

\bibitem{cohen}  Cohen, J.M., Gautreau, R.: Naked singularities, event horizons, and charged
particles, \PhR \textit{D} \textbf{19}, 2273-2279 (1979).

\bibitem{gladush}  Gladush, V.D.: The quasi-classical model of the spherical configurations in general
relativity, \IJMD  \textbf{11}, 367-390 (2002).

\bibitem{gonsal}  Gon\c{c}alves, S.M.: Shell crossing in generalixed Tolman-Bondi
spacetimes, \PhR \textit{D} \textbf{63},  124017-1-10 (2001).

\bibitem{wilkins}  Wilkins, D.: Bound Geodesics in the Kerr metric, \PhR \textit{D} \textbf{5}, 814-822 (1972).

\bibitem{felice} De Felice, F., Maeda, K.: Topology of collapce in conformal diagrams,
\textit{Prog. Theor. Phys.} \textbf{68}, 1967-1978 (1982).

\bibitem{bicak}  Bi\u{c}\'{a}k, J.,  Stuchl\'{i}k, Z., and  Balek, V.:  The motion of charged particles
 in the field of rotating charged black holes and naked singularities, \textit{Bull. Astron. Inst. Czech.}
\textbf{40}, 65-92 (1989).

\bibitem{finley}  Finley, J.D.:  Radial charged particle trajectories in the
extended Reissner-Nordstrom manifold, \JMP \textbf{15}, 1698-1701
(1974).

\bibitem{chandra}  Chandrasekhar, S.:  The mathematical theory of black
holes, 663 p.  Clarendon Press, Oxford (1983).

\bibitem{gron}  Gr{\o}n, O.: Repulsive gravitational and
 electron models,  \PhR \textit{D} \textbf{31}, 2129-2131 (1985).

\bibitem{cruz}  De la Cruz, V., Israel, W.: Gravitational Bounce,  \NCim \textit{I}
\textbf{A3}, 744-760 (1967).

\bibitem{leon} De Leon, J.P.: Gravitational repulsion in
sources of the Reissner-Nordstr\"{o}m field, \JMP \textbf{29},
197-206 (1988).

\bibitem{gorelik}  Gorelik, G.E.: Antigtavitation and electrical charge,
\textit{MSU Vestnik, series Fhys. Astron.} \textbf{13}, 727-728
(1972).

\bibitem{bonnor}  Bonnor, W.B.: The equilibrium of a charged test particle in the field of a
spherical charged mass in general relativity,  \CQG \textbf{10},
2077-2082 (1993).

\bibitem{landau}  Landau, L.D., Lifshitz, E.M.: The Classical Theory of
Fields, 402 p. Pergamon Press, Oxford (1975).

\bibitem{vikers}  Vickers, P.A.: Charged dust spheres in general relativity,  \textit{Ann. Inst. H.
Poincar\'{e}} \textbf{18}, 137-145 (1973).

\bibitem{graves}  Graves, J.C., Brill, D.: Oscillatory Character of Reissner-Nordstrom Metric for an Ideal
Charged Wormhole,  \PhR \textbf{120}, 1507-1513 (1960).

\bibitem{hawking} Hawking, S.W.,  Ellis, G.F.R.:   The Large Scale
Structure of Space-Time, 391 p. Cambridge Univ. Press, Cambridge,
England (1973).

\bibitem{frolov} Markov, M.A., Frolov, V.P.: On the minimal size of particles in the general theory of relativity, \TMF
\textbf{13}, 41-61 (1972).

\bibitem{glad-galad} Gladush, V.D., Galadgyi, M.V.: The radial motion of charged test particles and
 spherically symmetrical dust shells on General Relativity. In: Abstracts of 13 Russian Gravitational Conference
 --- International Conference on Gravitation, Cosmology and Astrophysics
(RUSGRAV-13), Moscow, PFUR, Russia, 31-32 (2008).

\bibitem{gladush07} Gladush, V.D.: On the stable spherically-symmetric charged dust
configurations in General Relativity,  \textit{Odessa Astron.
Publ.} \textbf{20}, 47-50 (2007).

\bibitem{gladush08} Gladush, V.D.:  The stable static spherically-symmetric charged dust configurations
in General Relativity. In: Abstracts of 13 Russian Gravitational
Conference --- International Conference on Gravitation, Cosmology
and Astrophysics (RUSGRAV-13), Moscow, PFUR, Russia, 30-31 (2008).

\end{thebibliography}


\end{document}